# "Frequency-modulated" pulsed Bell setup avoids post-selection.

Mónica Agüero, Alejandro Hnilo , Marcelo Kovalsky and Myriam Nonaka.
*CEILAP, Centro de Investigaciones en Láseres y Aplicaciones, UNIDEF (MINDEF-CONICET);*
*CITEDEF, J.B. de La Salle 4397, (1603) Villa Martelli, Argentina.*
email: ahnilo@citedef.gob.ar
July 5th, 2023.

Excepting event-ready setups, Bell experiments require post-selection of data to define coincidences. From the fundamental point of view, post-selection is a true "logical loophole". From the practical point of view, it implies a numerically heavy and time consuming task. In Quantum Key Distribution (QKD), it opens vulnerability in case of a hostile adversary. The core of the problem is to synchronize independent clocks during long observation runs. A pulsed source gets rid of clocks' drift, but there is still the problem of identifying the same pulse in each remote station. We use a frequency modulated pulsed source to achieve it. This immediately defines the condition of valid coincidences in a manner that is unaffected by the drift between the clocks. It allows finding the set of entangled pairs avoiding post-selection and in a way that is found to be optimal. It is also robust against a hostile adversary in the case of QKD.

**1. Introduction.**

Entangled states of photons are essential in experimental tests of Quantum Mechanics' (QM), in many processes involving quantum information, and in the practical application known as device-independent Quantum Key Distribution (QKD) [1]. However, filtering out the data to be included in the set of entangled pairs is not a trivial task, especially if the photons' detections are recorded in spatially distant stations. Having distant stations of observation is of interest in many cases, and is unavoidable in QKD. In these cases the observers get two lists (one for each station) of time values corresponding to photon detections, as measured by a local clock. The task is finding which time values, among all the recorded ones, correspond to coincident detections. This is known as *post-selection*. In order to perform it, the first step is to choose a time window value $T_w$. Time values in the lists of stations A and B such that $|t_A - t_B| \leq T_w$ belong to the set of coincident detections. But different distances, cable lengths and time response of instruments and detectors must be taken into account. The second step then is to add some *delay* time $d$ to one of the lists.

The way to determine the values of $T_w$ and $d$ is a combination of educated guess and iteration. One starts with a "large" value of $T_w$ and counts the total number of coincidences $N_c$ for different values of $d$. One scans a "reasonable" range of $d$ from estimated distances, cable lengths and instruments' response times. If the histogram of $N_c$ vs $d$ has a well defined maximum, this provides the first value of $d$. Then one can shorten $T_w$ and fine tuning $d$, in an iterative process. Each step in the iteration means a sort of convolution of one of the lists with the other. If the original time lists are long (as it usually happens) and the final value of $T_w$ is short (as it is usually aimed), post-selection is a numerically heavy and time consuming task. At the end, one gets a single peak of $N_c$ of width given by the shortest $T_w$ chosen. The set of selected coincidences is defined by the data under this peak.

Nevertheless, the iteration is not guaranteed to converge. The structure of the lists of time values can be intricate. More than one peak may appear at some point in the iterative process. Often the plots of $N_c$ vs $d$ have wide "platforms" around the main peak and display secondary peaks at large values of $d$. In the case the iterative process does not converge to a satisfactory histogram, then the usual interpretation is that something went wrong in that recording run and the data are discarded. If everything goes right instead, the data under the main peak allow calculating some parameter (say, $S_{CHSH}$) violating Bell's inequality. This means the set of data to correspond to an entangled state. Data in the secondary peaks sometimes violate the Bell's inequality too (although to a lesser extent than the ones in the main peak) and sometimes not. These features are believed to be caused by drift between the clocks in the remote stations. In one recorded case at least, the drift produced an effect important enough to allow an eavesdropper to break part of the key (if the set of data had been used for QKD) [2]. The Global Positioning System (GPS) is often used to synchronize the remote clocks.

Post-selection also has an edge from the point of view of the foundations of QM. According to the standard QM description [3,4] the pump field is written as a superposition of plane continuous monochromatic waves, so that the output state is the integral: $|\psi_{output}\rangle = \int d^3k \, |\psi[F(k)]\rangle$, where $|\psi[F(k)]\rangle$ is the state produced by the ideal plane wave with wave vector $k$ and amplitude $F(k)$. If the pump bandwidth $\Delta\omega_{pump}$ is smaller than the bandwidth of the spontaneous photon down conversion process in the crystals and also than the filters' bandwidth $\Delta\omega_{filters}$ (as it is usual) then the probability of detecting one photon of the entangled pair at time $t$ and the other photon at time $t'$ is:

$$P(t,t') \sim exp\, -\Delta\omega_{filters}^2 (t-t')^2 \times exp\, -\tfrac{1}{2}\Delta\omega_{pump}^2 (t+t')^2 \quad (1)$$

This means that the detections can only happen at times dictated by the pump, while their time correlation is defined by the filters' spectral width [3]. The shortest value for $T_w$ is given in practice by time jitter of the detectors ($\approx$2ns), which is much longer than the inverse of the bandwidths ($<$ 0,1ps). These numbers justify the intuitive picture (which underlies the procedure of post-selection) that entangled photons are like "bullets" that propagate from the source to the stations. Recall that, in the general case, it is not possible to define a wave function for the photon [5].

Note that a *logical* loophole arises: QM is assumed valid in order to derive eq.1. This equation and the involved bandwidth values validate the post-selection procedure. Data selected according to this procedure are used to test the validity of QM through the violation of Bell's inequality. But QM was assumed valid at start. J.S.Bell was aware of this loophole, that's why his early experimental proposals included an "event-ready" signal indicating that an entangled pair was emitted by the source [6], making post-selection unnecessary. Some event-ready Bell's experiments have been in fact performed [7-9]. They involve both photons and matter and the quantum phenomenon of entanglement swapping, which may lead to another logical loophole [10]. Anyway, they are very complex setups, difficult to use at the current state of technology.

In summary: from the foundational point of view, experiments free of the post-selection procedure are desirable. From the practical point of view, avoiding the sometimes ambiguous and always time consuming calculations of post-selection is also desirable. Using a pulsed source is a simple solution. Such a source is not "event ready", because it cannot certify an entangled state has been emitted. But it can certify when it has *not* been emitted. This approach has been used to successfully close the so called time-coincidence loophole [11]. A pulsed source also gets rid of clocks' drift through "logical synchronization" [12]. However, the problem of identifying the same pulse in both stations still remains. In this paper, we introduce the method of modulating the frequency of the pulsed source to fully circumvent post-selection. This method also dispenses with the GPS.

## 2. Setup.

The setup is sketched in Figure 1. Biphotons at 810 nm in the Bell state $|\varphi^+\rangle$ are produced in the standard configuration using two crossed BBO-I crystals and walk-off compensating crystals, pumped by a diode laser at 405 nm. This laser emits square pulses of 1 µs duration at a repetition rate of 500 Khz (50% duty cycle). These numbers are chosen so that the probability $p$ of detecting one photon during one pulse is $p<<1$. This condition is necessary to limit the number of accidental coincidences in the pulsed regime [13]. The entangled photons are inserted into single-mode optical fibers. Polarization is observed with fiber optic analyzers. Their relatively poor contrast (1:100) limits the value of $S_{CHSH}$ that can be achieved to 2.77. Silicon avalanche photodiodes detect single photons; time values of photon detection and trigger signal are stored in time-to-digital converters (TDCs). They have 10 ps nominal time resolution, but accuracy is reduced to ≈2 ns because of detectors' jitter. After pumping the crystals the laser beam is sent to a beam-splitter and illuminates two fast photodiodes. The resulting electrical signals are sent through 38m of coaxial cable to each station. Pulse shape distortion in this cable length is checked to be negligible. Two photodiodes are used (instead of just one) to avoid spurious echoes in the long cables. Three input channels are hence used in each TDC: two for the "1" and "0" outputs of the polarizer, and one for the trigger signal indicating the start of each pump pulse.

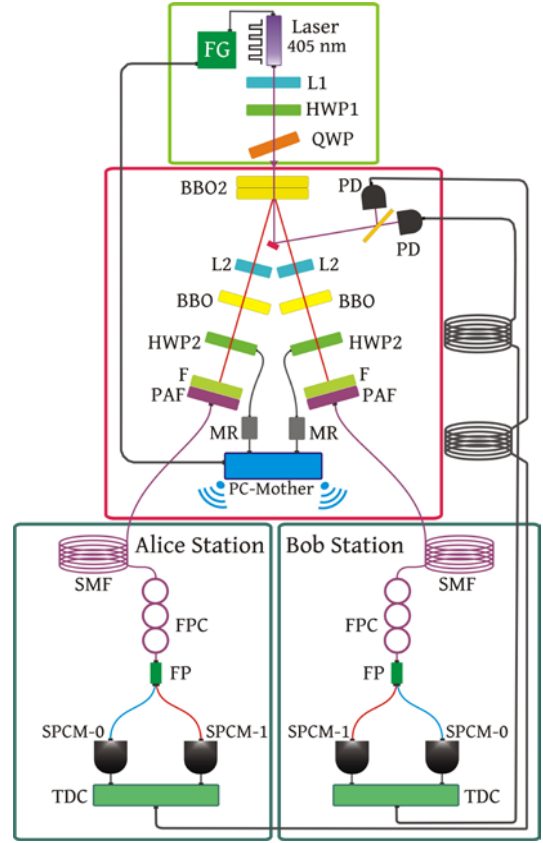

Figure 1: Sketch of the setup. FG: programmable function generator, L1,L2: focusing lenses, HWP1 and QWP: half and quarter waveplates at 405nm, BBO2: crossed BBO-I crystals, PD: fast photo-diodes, they record samples of the pumping pulse and send trigger signals to the TDCs through coaxial cables (38m long each), BBO: walk-off compensating crystals, HWP2: half-waveplates at 810nm to set the observation angles, MR: Rotating servo motors, F: filters at 810nm, Δλ=10nm, PAF: fiberports f = 7.5mm, SMF: single-mode fiber coils, FPC: birefringence compensators ("bat-ears"), FP: fiber polarization analyzers, SPCM: single photon detectors, TDC: time-to-digital converters. Stations are separated by 42m through the optical fibers and 2.64m in straight line. The setup is designed to allow the stations to be moved to remote places.

The optical fibers (21m long) and coaxial cables are currently coiled; they are placed in preparation to move the stations to distant positions. Note there is no link to synchronize the clocks other than the trigger signals sent through the coaxial cables. A "Mother" computer controls the function generator that pulses the laser and the servo motors that adjust the angle settings. She also instructs remotely through a TCP/IP communication via a local network, the "sons" computers in each station to open, name and close the data files recorded in each experimental run.

## 3. Tested method and results.

As said, synchronization between the clocks at the remote stations is a main issue. The pulsed regime eliminates the problem of the clocks' drift, for synchronization is refreshed with each pulse slope, and the drift is negligible during the typical pulse's duration. But, although Mother orders both sons to start recording data simultaneously, fluctuating differences between the time values each son actually starts to record are observed to be as large as 10 ms, what means a difference of $\approx 5\times 10^4$ pumping pulses. In these conditions, the only way to identify the same pulse (among the $5\times 10^4$ ones) in both stations is by counting coincidences, that is, to perform the unwanted post-selection procedure. We find a simple alternative by modulating the frequency of the pump pulses, as it is explained next.

Before a recording run starts, the pulse frequency is set to 490 Khz (i.e., slightly different from the value used during actual recording). When everything is ready to start the run, Mother orders the generator to switch to 500 Khz. This produces a sharp step in the pulse separation without affecting the pulse shape. That step is easily identified in the trigger files in each TDC, see Figure 2. This determines the "first pulse in the run" in both stations and allows numbering the following pulses. Once the run (10 to 30s in real time) has ended, Mother switches the pulsing frequency back to 490 Khz and orders the sons to save the recorded files and to prepare for the next run.

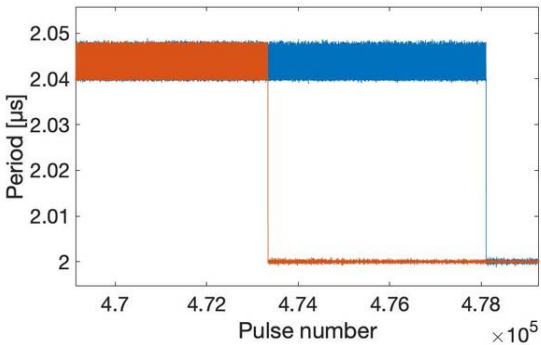

Figure 2: Illustration of the method to identify the same pulse in each station. The period of the train of pulses is shown as observed in the Alice (Bob) station in red (blue). As the step is produced by the same event, numbering the following pulses is immediate. No post-selection is necessary.

In the run in Fig.2, the time difference between the moment Alice and Bob start measuring is found by the difference between the locally measured pulse numbers where the step occurs (478113 – 473343) multiplied by the period (2 µs), that is, 9.54 ms. This time is different in each run, but is easily found in this way. As $p<<1$, if a detection occurs in pulses with the same number in each station, then a coincidence is immediately found. This criterion is unambiguous, fast and reliable. Further filtering could be done by using $T_w$ shorter than the pulse duration, but is found unnecessary (see below).

In our case, a single switch event at start suffices to number all pulses in the run. In runs longer than 30s, the TDCs may fail to record one or more trigger signals during the run. In this case, the function generator can be easily programmed to modulate the pulsing frequency (say, by introducing a slow chirp). This allows identifying the missing trigger signals and restoring pulse numbering.

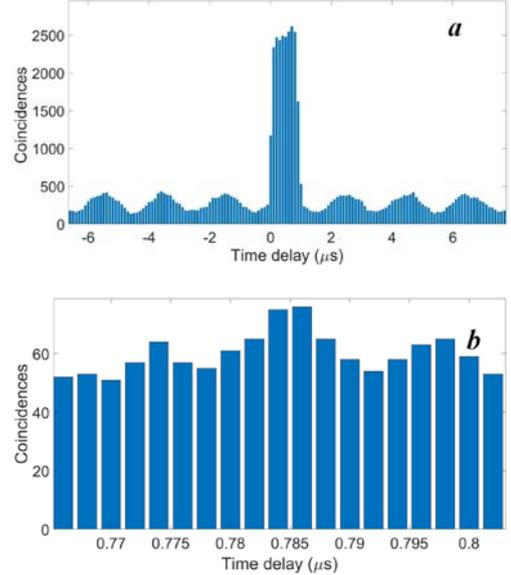

Figure 3: Histograms of $N_c$ vs $d$ obtained by post-selection, (a) $T_w = 100$ ns, (b) $T_w = 2$ ns.

As an illustration, a typical histogram obtained during post-selection is displayed in the Figure 3a. The relatively broad main peak and the secondary peaks are caused by the drift between the independent clocks. There are no ambiguities in this case; anyway, obtaining this histogram means performing a heavy numerical task. In the Figure 3b, the result of carrying post-selection further down to $T_w = 2$ ns is shown. Note the scarce statistics and the difficulty in defining the main peak. The value of $S_{CHSH}$ calculated with $T_w = 2$ ns is hence meaningless. With these data, $S_{CHSH}$ must be calculated with $T_w \geq 100$ ns.

In the method we test in this paper, drawing histograms like the ones in Figs.3 is unnecessary. In the Figure 4, instead, we show the histogram of the time distances between the detections (summed up during the complete run) observed in Alice and Bob *in the pulses with the same numbering*. As it is seen, filtering with $T_w$ shorter than the pulse duration is redundant.

The original (experimentally recorded) raw data files are the same than in Figs.3 and 4. They correspond to setting $\alpha=0$, $\beta=0$, and "1" outputs in both stations in Fig.1; recording time $\approx 10$ s, number of pulses 4,511,169; $N^A_{singles} = 115,861$; $N^B_{singles} = 108,874$

($\Rightarrow p \approx 0.024 \ll 1$ as required). For this file $N_c = 2,614$ by using post-selection with $T_w = 100$ ns (Fig.3a) and $N_c = 8,794$ by using our method with $T_w = 2$ ns (Fig.4). Note the huge difference with Fig.3b in the number of coincidences (and hence, in the efficiency).

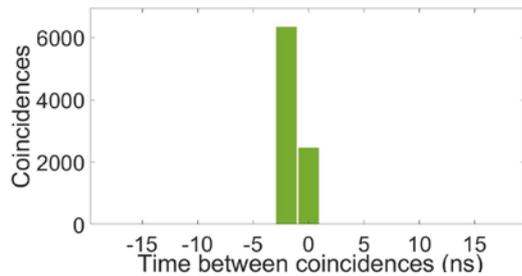

Figure 4: Histogram of the time distance between detections in Alice and Bob that occur during the same pulse, the original raw data files are the same than in Fig.3. The time resolution is 2 ns.

Similar files recorded with additional settings allow calculating the level of entanglement. Using always the same raw experimental files, coincidences obtained by post-selection lead to $S_{CHSH} = 2.14 \pm 0.11$ ($T_w = 100$ns). The ones obtained by our method lead to $S_{CHSH} = 2.78 \pm 0.05$ ($T_w = 2$ns), reaching the limit that is possible with our fiber polarization analyzers. Therefore, in addition to the faster and unambiguous processing of data, the entanglement of the selected set is higher.

## 4. Summary and conclusions.

This paper essentially deals with the problem of synchronization between remote clocks in a Bell setup. This is of interest for experiments on QM foundations and for QKD. A customary solution is using the signals provided by the GPS. But, in the case of QKD, it has the drawback that the GPS can be jammed or destroyed by a hostile adversary, making communication unreliable or impossible.

The tested method dispenses with the existence of the GPS. Its key features are the pulsed regime and the modulation of the repetition rate. The first feature gets rid of clocks' drift, the second one allows pulse numbering even if one or more pulses fail to be detected by the TDCs. Pulse numbering is found to suffice to determine coincidences, further filtering is redundant.

In conclusion, the tested method allows finding the set of entangled pairs between remote stations in a fast, unambiguous and efficient way. Besides, it is free of the logical loophole of post-selection. We believe the tested method to be of wide interest in both pure and applied experimental research in Quantum Information.

**Acknowledgments.**

This work received support from the grants N62909-18-1-2021 Office of Naval Research Global (USA), and PUE 229-2018-0100018CO CONICET (Argentina).